\documentstyle[prl,aps,multicol,epsfig]{revtex}
\begin{document}
\draft
\title{  Theory for Dynamical Short Range Order 
and Fermi\\ 
Surface Volume in Strongly Correlated   Systems}

\author{J. Schmalian, M. Langer,  S. Grabowski, 
               and K.H. Bennemann}
\address{Institut f\"ur Theoretische Physik,
  Freie Universit\"at Berlin, Arnimallee 14, \\
       14195 Berlin, Germany}
\date{ March 26, 1996}
\maketitle 
\begin{abstract} 
\leftskip 54.8pt
\rightskip 54.8pt
Using the  fluctuation 
exchange approximation of the one band
Hubbard model, we discuss   the origin of the 
 changing   Fermi surface volume 
in underdoped cuprate systems due to the transfer 
of occupied states from the Fermi surface
to its  shadow, resulting 
  from the strong dynamical 
antiferromagnetic short range correlations.
The momentum and temperature dependence of
the quasi particle  scattering rate shows unusual
deviations from the  conventional  Fermi liquid like behavior.
Their consequences   for the changing Fermi surface volume 
 are discussed. Here, we  investigate in detail  which 
scattering processes might be responsible for  a violation
of the Luttinger theorem.
Finally, we discuss   the formation of hole pockets near half filling.
 \end{abstract}   

\begin{multicols}{2}   

\narrowtext
%
%
%
It is clear that the evolution of  the Fermi surface (FS)
upon doping and 
the occurrence of a very flat band 
crossing the Fermi level
 is of great importance  for an understanding 
of the anomalous properties
of high-T$_c$ materials.~\cite{KS90,BSW94,A94,DNB94,haas,PHvdL95,LSG95,CMS95}
 Near half filling, the cuprates are antiferromagnetically 
ordered.
Therefore, one expects due to the low carrier concentration 
small Fermi surface pockets. ~\cite{T88DJM90}
Strong indications for this behavior were experimentally 
observed in  YBa$_2$Cu$_3$O$_{6.3}$~\cite{LVP92} 
and in Sr$_2$CuO$_2$Cl$_2$~\cite{WSM95}.
In contrast, for optimally and overdoped cuprates, 
large Fermi surfaces  are well
established.~\cite{SD95} 
One could simply expect a transition from a small to 
a large FS at the antiferromagnetic-paramagnetic
phase transition, occurring for $x \approx 0.025$ in 
La$_{2-x}$Sr$_x$CuO$_4$.
Interestingly, no abrupt changes of the transport 
properties or the details of the electronic
structure were  found  at this 
transition.~\cite{Rom90,KBB92}
 Furthermore, pronounced short ranged 
antiferromagnetic correlations were observed in the
paramagnetic state, even for optimally doped 
materials.~\cite{KBB92,BGJ88,ISY93}
An alternative scenario for the transition from a large 
to a small FS is the  gradual evolution of shadows 
of the FS in the paramagnetic 
state~\cite{KS90,haas,LSG95,CMS95}, first observed 
by Aebi {\em et al.}~\cite{AOS94}.
In a recent letter~\cite{LSG95}, we gave a  
theoretical explanation for these 
   states  with correct excitation energy and 
consistent with the observed relatively short antiferromagnetic 
correlation length. Here, the shadow states reflect a dynamically 
broken symmetry  on a short range  which demonstrates  the dynamical
 character of these new structures in the excitation 
spectrum of the cuprates.~\cite{LSG96}
This is related to an unusual behavior of the
electronic self energy leading  for low doping concentrations 
to  a small  change of the Fermi surface volume
compared  to the uncorrelated case for equal particle
density.

It is the aim of this paper to  discuss  in detail 
the anomalous momentum and temperature 
dependence of the quasi particle scattering rate and  how
it causes  the  change of the Fermi surface volume
of  underdoped   systems due to the 
occurrence of shadow states outside the Fermi
surface.
Although quantitatively small, the change of the 
Fermi surface volume, found in our FLEX calculations~\cite{LSG95},
 are  directly related to these new structures
in the spectral density.
It will be shown that the anomalous frequency, 
momentum and  temperature dependence
of the  electronic self energy, reflecting the 
dynamical character of the antiferromagnetic
correlations,   are responsible for this qualitatively 
new behavior.  
Furthermore, we  argue that the dynamical formation of
short range correlations which modifies the translation
symmetry of the underlying lattice on a certain
time scale is a general phenomenon related to  the  non
Fermi liquid behavior of  a strongly correlated system.
Therefore, the original derivation of the  Luttinger theorem~\cite{LW60,L61} 
is  extended to an arbitrary frequency 
and momentum dependence of the quasi
 particle scattering rate, not restricted to the 
usual  Fermi liquid behavior 
${\rm Im}\Sigma_{\bf k}(\omega) \propto \omega^2$.
This enables us to investigate in detail  
which scattering processes are responsible for  a violation
of the Luttinger theorem.
This  might be of   importance for an 
understanding of the evolution of the Fermi surface 
of the high-T$_c$ materials upon doping.
Finally, we discuss that the   changing FS-volume
and the formation of hole pockets near half filling    are
both closely related with the formation of shadow states.

The  shadow states derived within the self consistent and 
conserving fluctuation exchange approximation
(FLEX)~\cite{BS89,BSW89} by Langer {\em et al.}~\cite{LSG95}
result  from a dynamical coupling of states with momentum 
${\bf k}$  at the FS with   states  at its shadow at ${\bf k}+{\bf Q}$,
  where $ {\bf Q}=(\pi,\pi)$.
Due to the  absence of long range antiferromagnetic order,
this coupling has to be 
 generated via   decay of particles with momentum ${\bf k}$
into a state with ${\bf k}+ {\bf Q}$ and vice versa.
This is related to an unconventional behavior of the quasi particle
scattering rate, i.e. of the imaginary part of the self energy
$\Sigma_{\bf k}(\omega)$.
By calculating $\Sigma_{\bf k}(\omega)$ within the FLEX approximation 
  of the one band  Hubbard model~\cite{SLG96}, we found this
 anomalous momentum and frequency  dependence.

\begin{figure}
\centerline{\epsfig{file=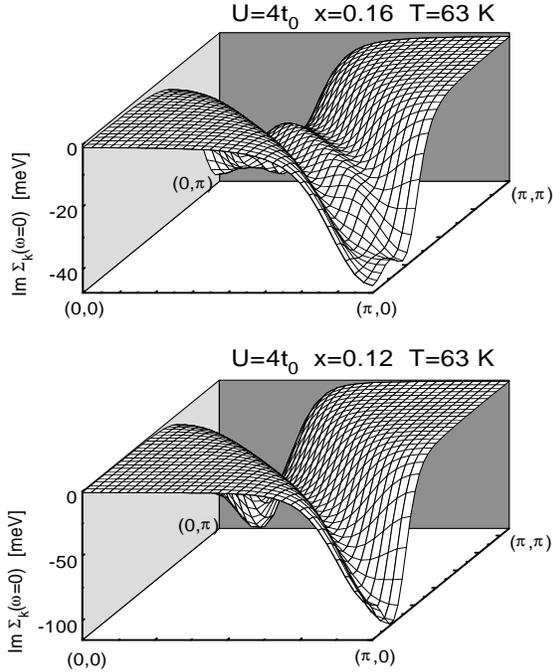,width=9cm,height=11cm}}
\caption{ Momentum-dependence of   ${\rm Im} \Sigma_{\bf k}(0)$
   demonstrating the strong increase of quasi
particle scattering on the shadow of the Fermi surface
for decreasing doping. }
\label{fig1}
\end{figure}

In Fig.~\ref{fig1}, we show our results for the momentum dependence of 
${\rm Im}\Sigma_{\bf k}( 0)$ for two different doping concentrations $x=1-n$.
Here, $n$ is the particle density per lattice site.
The calculation was performed using an
unperturbed dispersion 
$\varepsilon_{\bf k}=-2t(\cos(k_x)+\cos(k_y))$ 
with nearest neighbor hopping element $t=0.25\, {\rm eV}$,
and  a value of the on site Coulomb interaction  $U=4t$. 
While for larger doping ($x=0.16$) a double well structure with large quasi particle
scattering rate at the Fermi energy and at its shadow occurs, we find for $x=0.12$ 
the dominant scattering phenomena  are at the FS-shadow, leading to the formation
of shadow states in the spectral density.

\begin{figure}
 \centerline{\epsfig{file=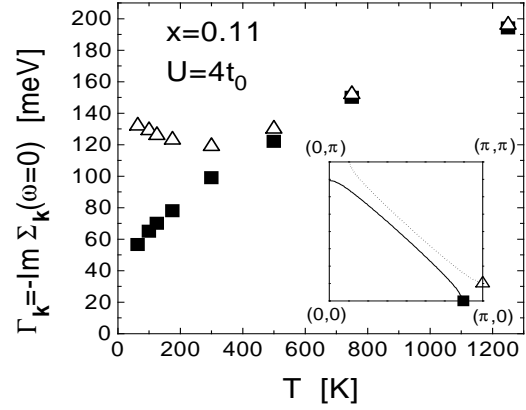,width=9cm,height=11cm}}
\vskip -4.8cm
\caption{ Temperature dependence of the
 quasiparticle scattering rate
$\Gamma_{\bf k}=-{\rm Im}\Sigma_{\bf k}
(\omega=0)$ for a ${\bf k}$-value
on the Fermi surface (solid squares) and on 
the shadow of the Fermi surface 
(open triangles), as indicated in the inset.
Note the anomalous increase of  
$\Gamma_{\bf k}$ on the shadow 
for decreasing temperature.}
\label{fig2}
\end{figure}

In Fig.~\ref{fig2} we show the temperature dependence of 
 $\Gamma_{\bf k}\equiv  -{\rm Im}\Sigma_{\bf k}(\omega=0)$
for ${\bf k}$ at the Fermi surface (solid squares) and at its shadow (open triangles),
which dramatically differs from
the conventional $\Gamma_{\bf k} \propto T^2 $ dependence
occurring in  overdoped systems.
The scattering rate at the FS-shadow increases with decreasing temperature
since  the corresponding short range correlations, 
which build up the ${\bf k}$-dependencies
of $\Gamma_{\bf k}$,  are getting stronger for $T \rightarrow 0$.

 In view of this anomalous behavior of the electronic self energy, it is of 
interest to explore its consequences on the shape and volume
of the Fermi surface.
Therefore, we consider the 
 FS-volume, characterized
by
\begin{equation}
n_{\rm Lutt.}(T)=\frac{2}{N} \sum_{\bf k} \theta \left( 
\mu - \varepsilon_{\bf k} -
{\rm Re}\Sigma_{\bf k}(\omega=0) \right)\, ,
\end{equation}
and compare it with the particle concentration
\begin{equation}
n(T)= -\frac{2}{ N}
\sum_{\bf k} \int_{-\infty}^{\infty} \frac{d \omega}{\pi}
f(\omega) {\rm Im} \, G_{\bf k}(\omega + i0^+) \, .
\end{equation}
Here, $G_{\bf k}(z)=(z+ \mu - \varepsilon_{\bf k}-
 \Sigma_{\bf k}(z))^{-1}$ is the Green's function with
chemical potential $\mu$,  
and self energy $\Sigma_{\bf k}(z)$, respectively.
 $f(\omega)=(e^{\omega/T}+1)^{-1}$ is the Fermi function, and
$N$ is the number of lattice sites.

\begin{figure}
\vskip -1.5cm
\centerline{\epsfig{file=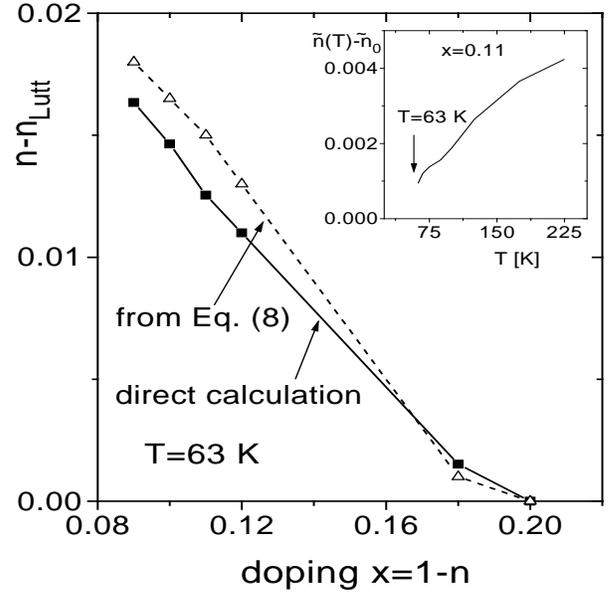,width=9cm,height=10cm}}
\caption{Doping dependence of $n-n_{\rm Lutt.}$ , with 
  particle density $n$ and
  volume of the Fermi surface $n_{\rm Lutt.}$.
For low doping concentration, a  change of the
FS-volume compared to $U=0$
occurs.  
The agreement between  the solid and dashed line
demonstrates the applicability of Eq. 8 which relates
the ${\bf k}$-dependence of the scattering rate with the
FS-volume.
The results refer to a temperature
 $T=63 \, K$ and a value of 
the Coulomb repulsion  $U=4t$ with 
nearest neighbor hopping 
element $t=0.25\, {\rm eV}$.
The inset shows the temperature dependence of 
$\tilde{n}(T)-\tilde{n}_o$, which is smaller than 
the difference $n-n_{\rm Lutt.}$ 
showing that the  influence of Fermi function smearing  
on the violation of the LT can be 
neglected in our finite temperature 
calculation.} 
\label{fig3}
\end{figure}

 We show in Fig.~\ref{fig3} our  results  for
the doping dependence
of the   difference $n-n_{\rm Lutt.}$ (solid line)   demonstrating
that in underdoped systems the volume 
of the Fermi surface  differs from its value $n$   for $U=0$.
This is of importance since  
\begin{equation}
n(T)=n_{\rm Lutt.} (T)
\label{LT}
\end{equation}
is for $T=0$ the Luttinger theorem~\cite{LW60,L61} (LT), implying an
independence of the FS-volume on the interaction strength
for given particle density.
 We argue on physical grounds that this shrinking
of the Fermi surface~\cite{FN1} results from a 
transfer of occupied states from the main FS
to its shadow~\cite{LSG95} and gives strong 
indications for a violation of the LT, 
 and is not due to the finite temperature of the calculation.
The following analysis of our numerical results  
strongly supports  this point of view although
  such a transfer of   states does not
 necessarily leads to a violation of the LT,
since $n_{\bf k} <1$ inside and $n_{\bf k} > 0$ outside  
the FS is   well known  for
interacting  fermions.  

For an explanation of the deviation of $n_{ \rm Lutt.}(T)$ from
 $n(T)$, it is helpful to reconsider the   derivation of the Luttinger theorem.
 The Luttinger theorem applies if: (i) perturbation theory is applicable, 
(ii) the theory is   conserving~\cite{BK61},
(iii) the temperature is zero, 
and (iv) the imaginary part of the self energy  
vanishes at the Fermi energy, e.g. like
${\rm Im}\Sigma_{\bf k}(\omega) \propto \omega^2$.
Note that for an arbitrary FS-shape (iv) does not follow from the first
three prerequisites.
In the following we  investigate in particular the  importance of  (iii) and (iv)  on
a possible violation of the LT.
 We  start from the 
following identity for the particle density 
valid for arbitrary temperatures:
\begin{equation}
n(T)=\frac{2 T}{N} \sum_{{\bf k}, m} 
G_{\bf k}(i \omega_m) e^{i \omega_m 0^+}=
 \tilde{n}(T) + I(T)\, ,
\label{decomp}
\end{equation}
with 
\begin{equation}
\tilde{n}(T)=\frac{2 T}{N} \sum_{{\bf k}, m} 
\frac{\partial}{\partial(i \omega_m) }{\rm ln} \left(
G_{\bf k}(i \omega_m) ^{-1} \right) 
e^{i \omega_m 0^+}
\label{ntilde1}
\end{equation}
and 
\begin{equation}
 I(T)=\frac{2 T}{N} \sum_{{\bf k}, m}
G_{\bf k}(i \omega_m )\frac{\partial
 \Sigma_{\bf k}(i \omega_m)}
{\partial(i \omega_m)}
e^{i \omega_m 0^+}\, ,
\label{I(T)}
\end{equation}
which follows immediately if one takes 
Dyson equation into account.
 Here, $\omega_m=(2m+1)\pi T$
is a fermionic Matsubara frequency. 
 As   shown by Luttinger and 
Ward~\cite{LW60} $I(T)$ vanishes for 
zero temperature  order
by order in perturbation theory if the
 theory is conserving,
i.e. if there exists a diagramatically well defined
 functional $\Phi[G]$ which generates the self energy
via functional derivation~\cite{BK61}.
Since our numerical data are obtained for 
finite temperature, we  obtain a finite value 
$I(T) \approx 10^{-3}$,  which nevertheless can
be neglected  compared to $n-n_{\rm Lutt.} \approx 10^{-2}$.
Considering now $\tilde{n}(T)$ in Eq.~\ref{ntilde1},
analytical continuation and partial integration yield
\begin{eqnarray}
\tilde{n}(T)&=&-\frac{2 }{N T} \sum_{{\bf k}} 
\int_{-\infty}^{\infty} \frac{d \omega}{\pi} 
{\rm Im} \left[ {\rm ln} \left( 
G_{\bf k}(\omega+i0^+ ) ^{-1} \right) \right] \nonumber \\
& & \times f(\omega)\left(1-f(\omega)\right)\, .
\label{ntilde2}
\end{eqnarray}

In order to show that our conclusion concerning 
the violation of the LT is also not influenced by some
Fermi function smearing due to finite temperatures,
we compare $\tilde{n}(T)$  with  $\tilde{n}_o$ which
is calculated from Eq.~\ref{ntilde2}  using the zero temperature
limit of $f(\omega)(1-f(\omega))/T \approx \delta(\omega)$. 
Note that $\tilde{n}_o \neq \tilde{n}(T=0)$ due to
 the additional temperature dependence of the
self energy $\Sigma_{\bf k}(0)$.
In the inset of  Fig.~\ref{fig3}, we show our data for 
the temperature dependence of
$\tilde{n}(T)-\tilde{n}_o$,  
 showing that for low temperatures $\tilde{n}(T)-\tilde{n}_o\approx 10^{-3}$ is  
  much smaller than $n-n_{\rm Lutt.}$, 
  leading to  
$n  \approx \tilde{n}_o$ with very good accuracy.
Thus, 
 \begin{equation}
n \approx n_{\rm Lutt.}+ \frac{2}{\pi N} \sum_{\bf k}
 {\rm arctan}\left(\frac{\Gamma_{\bf k}}{\varepsilon_{\bf k}
+{\rm Re}\Sigma_{\bf k}(0)-\mu}\right)\, .
\label{LT2}
\end{equation}
Eq.~\ref{LT2} is exact in the limit $T \rightarrow 0$ 
and can then be considered as a generalization 
of Luttingers original derivation without the assumption 
 ${\rm Im}\Sigma_{\bf k}(\omega=0) =0$,
while keeping the  prerequisites (i)--(iii) 
discussed above. 
Now, the volume of the FS is in general no more
 independent of the interaction strength.
A   violation of the LT occurs if the excitations 
at the Fermi surface cannot be described as ideal
quasi particles with  ${\rm Im}\Sigma_{\bf k}(\omega=0) =0$.
 Note that  one can separate in Eq.~\ref{LT2} 
one conventional  contribution $n_{\rm Lutt.}$, independent
of the  scattering rate,  from a new contribution 
which strongly depends on the details of $\Gamma_{\bf k}$.
Consequently, the {\bf k}-dependence of 
$\Gamma_{\bf k}$ gives a lot of interesting insights 
into the the origin of the difference 
$n-n_{\rm Lutt.}$.
If $\Gamma_{\bf k} \neq 0 $ symmetrically  
for the states insight and outside the FS, the
corresponding  contributions from the 
second term in Eq.~\ref{LT2} 
 cancel each other and $n=n_{\rm Lutt.}$ is fulfilled.
A small value for $n-n_{\rm Lutt.}$ is also expected for systems 
with momentum independent self energy and large
Fermi surface, independent on the details of the
scattering mechanism.
If however $\Gamma_{\bf k} \neq 0 $ predominantely
 for ${\bf k}$-states outside (inside) the FS, it
follows from  Eq.~\ref{LT2} that $n-n_{\rm Lutt.}>0$ 
($<0$), i.e. the FS volume shrinks (gets larger)
compared with the uncorrelated system.
Therefore, a strong ${\bf k}$-dependence of $\Gamma_{\bf k}$
different for states inside and outside the FS is a general
phenomenon leading to a change of the FS-volume.
Consequently, the dynamical excitations which lead to 
$\Gamma_{\bf k} \neq 0$ should be related to some short range order 
which modifies the translational symmetry and thus results in a 
corresponding strong momentum dependence.

Using our self consistently determined results for
the electronic self energy for the lowest temperature
$T=63\, {\rm K}$, the change of the FS-volume  can
indeed be described with Eq.~\ref{LT2} which is
fulfilled with the expected accuracy $\approx 10^{-3}$
for all doping concentrations, as shown
in Fig.~\ref{fig3} (dashed line).
Following the above argumentation, the FS shrinks,
since $\Gamma_{\bf k}$ is largest   on the shadow of the 
Fermi surface outside the main FS (see Fig.~\ref{fig1}
and Fig.~\ref{fig2}).  
 Therefore, the change of the FS-volume is
a consequence of the transfer of
occupied states to the shadow of the FS
due to a dynamical coupling of FS-states  
with states on the shadow
of the FS,
which confirms our original physical 
argumentation of Ref.~\onlinecite{LSG95}.
Such a coupling of single particle states is directly related
to a violation of the Fermi liquid theory, since
its basic idea is the one to one correspondence
of the quasi particles of the interacting
system with that of an ideal Fermi gas which
is no more guaranteed if it is impossible to
distinguish reasonably between the states
${\bf k}$ and ${\bf k}+{\bf Q}$.
Consequently, the   phase space arguments
of Landau's Fermi liquid theory are no more applicable.
For larger doping, the $T$-dependence of
 $\Gamma_{\bf k}$ indicates strongly
 $\Gamma_{\bf k}=0$ for $T \rightarrow 0$.
The shadow states  disappear and we find
$n=n_{\rm Lutt.}$, i.e. LT is fulfilled.

 Now we investigate how our numerical 
finite temperature results for $\Gamma_{\bf k}$
 might continue in the limit $T \rightarrow 0$
 which was performed to obtain Eq.~\ref{LT}.
This is of  importance since    the temperature 
dependence of $\Gamma_{\bf k}$,  shown 
in Fig.~\ref{fig2},
strongly suggests that $\lim_{T \to 0} 
\Gamma_{\bf k} \neq 0$, 
at least for the ${\bf k}$-values on the 
shadow of the FS.
Considering the FLEX equations in the zero
temperature limit,  the contribution $V^{(s)}_{\bf q}(\omega)$  
of the effective interaction~\cite{SLG96} which
results from the  longitudinal and transversal spin
fluctuations
behaves for low frequencies like
\begin{equation}
V^{(s)}_{\bf q}(\omega)= \frac{3  }{2} \frac{ U^2 \chi_{\bf q}(\omega)}{1-U\chi_{\bf q}(\omega)} 
\approx \frac{3  }{2} \frac{c_{\bf q}}{\omega^*_{\bf q}-i\omega} \, ,
\label{VT0}
\end{equation}
with   characteristic  spin excitation
 energy  
$\omega^*_{\bf q}=c_{\bf q}(1-U{\rm Re}
 \chi_{\bf q}(0)) / U $ and 
$c_{\bf q}= (\partial {\rm Im} \chi_{\bf q}(\omega)/
\partial \omega)|_{\omega=0})^{-1} >0$.
Here we  assumed a conventional low frequency behavior for the particle hole bubble 
$\chi_{\bf q}(\omega) \approx U^{-1}-  ( \omega^*_{\bf q}-i \omega)/c_{\bf q}.$
 It is important to note that
 $\omega^*_{\bf q}$ should not be
 confused with the
characteristic     excitation energy in 
the  dynamical  spin susceptibility.
The difference between the effective 
interaction of the FLEX and the  spin
susceptibility, where  in such an extreme limit
vertex corrections have to be considered, 
 was recently demonstrated  in Ref.~\onlinecite{SH95}.
We expect that even if $\omega^*_{\bf q}=0$ the 
coupling of the various spin modes due
to short range correlations  
leads to a finite excitation energy in the spin 
susceptibility, in agreement with 
the experiment for doped cuprates.
The evaluation of the FLEX diagrams yields
$\Gamma_{\bf q} \neq 0$ only if 
$\omega^*_{\bf q}=0$ for a 
momentum ${\bf q} \approx {\bf Q}$.  
Then it follows
\begin{equation}
\Gamma_{\bf k}=\frac{3}{2} \sum_{\bf q}'
 \varrho_{{\bf k}-{\bf q}}(0)\, c_{\bf q}\, .
\label{GT0}
\end{equation}
As expected from Fig.~\ref{fig1}b, $\Gamma_{\bf k}$ is  maximal on the
shadow of the FS,
since the spectral density at the Fermi level 
$\varrho_{{\bf k} }(0)$  is largest  for
${\bf k}$  on the Fermi surface and because 
the summation is performed only for 
${\bf q} \approx {\bf Q}$ (indicated by the 
prime in Eq.~\ref{GT0}). 
Our finite $T$ calculation   
shows that $\omega^*_{\bf q}$ gets extremely
small for ${\bf q} \approx {\bf Q}$ 
(we find $\omega^*_{\bf Q} \approx 
1-5 \, {\rm meV}$
which is   our numerical 
resolution)
and decreases for decreasing $T$.
This refers to a quasi instantaneous coupling
of states at the FS and its shadow such that a large amount
of scattering processes occurs already for very small excitation
energies or temperature.
Nevertheless,  we cannot say 
whether $\omega^*_{\bf q} 
\rightarrow 0$ occurs strictly
for $T \rightarrow 0$.
If $\omega^*_{\bf q}$ remains finite 
the scattering rate would vanish 
for very small temperatures  and 
our argumentation concerning the LT refers to
 $T > \omega^*_{\bf q}$.
 Note,  $\omega^*_{\bf q}=0$  is 
 not necessarily  related to a long
range ordered state.  
The short range order leads to a mode mode 
coupling resulting in a finite but nonsingular scattering rate.
Therefore, it is possible and  physically reasonable that our
 numerical results for finite $T$ are representative
for the   behavior at zero temperature.
In any case, our conclusion concerning the violation of the LT 
is  valid with good approximation for 
finite but very small temperatures $T > \omega^*_{\bf q}$.

\begin{figure}
\centerline{\epsfig{file=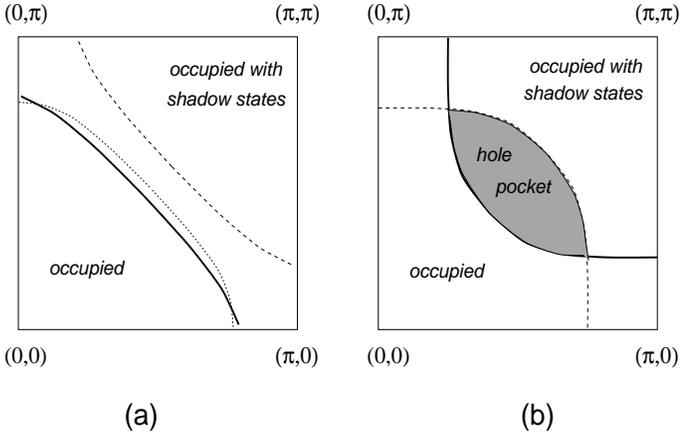,width=9cm,height=5.8cm}}
\vskip 1cm
\caption{ (a) main Fermi surface (solid line),  shadow of the Fermi surface (dashed line),
and  Fermi surface for $U=0$ (dotted line),
 and (b)  for the motion of   holes
within an antiferromagnetic background 
near half filling.
While the change of the  main FS-volume is rather small, 
in both cases additional states outside the FS are
occupied.
This  leads near half filling to the formation of hole 
pockets around $(\pi/2,\pi/2)$  (shadded area).}
\vskip 25cm
\label{fig4}
\end{figure}

Finally, it is of interest to contrast our quantitatively small
FS-changes with the formation of hole pockets, which occur
 if one considers the motion of holes within an 
antiferromagnetic background, relevant near half filling~\cite{T88DJM90}.
 As  indicated in Fig.~\ref{fig4}a, the formation of shadow states
 leads to a considerable
additional occupation outside the main Fermi surface.
Due to particle conservation this results from a decrease
of $n_{\bf k}$ inside the FS and also from the slight shrinking
of its volume discussed in this paper.
Consequently, the only unoccupied region of the Briouilline zone
is near the diagonal from $(0,\pi)$ to $(\pi,0)$. 
Since strong coupling calculations near half filling yield a pronounced
effective intrasublattice hopping~\cite{T88DJM90}, additional deformations of
the Fermi surface shape are expeced to occur for even smaller doping.
 This is schematically indicated in  
 Fig.~\ref{fig4}b. Now, the corresponding unoccupied region of the  Brillouin zone
is the hole pocket around the $(\pi/2,\pi/2)$ point.
A similar shape of the FS, and correspondingly of its shadow, occurs also if
one takes from the beginning an additional next nearest neighbor
hopping in $\varepsilon_{\bf k}$ into account, which  is believed
to be of importance for Bi$_{2}$Sr$_{2}$CaCu$_{2}$O$_{8+\delta}$ and
YBa$_{2}$Cu$_{3}$O$_{6+\delta}$~\cite{radtke}.
Thus, we  conclude that the changing FS-volume discussed in this paper and
the hole pocket formation are   of  common origin:
the transfer of states from the main to the shadow Fermi surface.

In conclusion, we discussed the consequences of an anomalous momentum
and temperature dependency of the quasi particle scattering rate for the
volume of the Fermi surface. Therefore, we extended the calculation of 
Luttinger and Ward~\cite{LW60} 
to the case of a finite
 quasi particle scattering rate $\Gamma_{\bf k}$.
 A non Fermi liquid  behavior results in a 
violation of the LT if the ${\bf k}$-states
with $\Gamma_{\bf k} \neq 0$ are asymetrically 
distributed with respect to the Fermi surface.
Furthermore, we showed that our numerical
 results obtained from the solution of the 
FLEX equation give evidence that the difference 
$n-n_{\rm Lutt.}$ obtained for small but
 finite temperatures reflects 
indeed a violation of the LT in underdoped cuprates
and are not caused by the finite $T$ of the calculation.
This  conclusion requires   that
$\Gamma_{\bf k}$ shown in Fig.~\ref{fig2} remains
 finite also for $T \rightarrow 0$.
Based on the zero temperature behavior of
the FLEX, we argued that this is  at least 
consistent with a state of strong dynamical
short range order such that our numerical results
might be representative also for $T=0$.
In any case, the anomalous behavior of the
scattering rate occurs for temperatures larger
than an extremely small energy scale $\omega_{\bf q}^*$
and shows that dynamical
short range correlations change dramatically the
single particle properties of the cuprates for  
physically relevant finite temperatures.
The origin of this interesting behavior 
of $\Gamma_{\bf k}$ is the occurrence of a 
many particle mode  leading to  a coupling of
 FS-states with states on the shadow of the FS.
Furthermore, we discussed  that 
the relatively small changes of the FS-volume and
the formation of hole pockets are of
 common origin.

Finally, we believe that the result of
 Eq.~\ref{LT2} might also be stimulating for 
the characterization of the
details of the Fermi surface in correlated systems, 
where the non  Fermi liquid  character is due to a 
totally different origin than that discussed in this paper.
One example  could be the  formation of a striped 
phase in La$_{2-x}$Sr$_x$NiO$_4$,~\cite{TBS94} where  
a modification of the translation invariance perpendicular
to the stripes occurs which seems to   result from 
 the strong dynamical character of the
corresponding excitations~\cite{EGS96}.
Other examples might be dynamical charge modulations,
i.e. dynamical CDW-states or the formation bipolarons
above its condensation temperature.
In all these cases we expect, similar to 
the case where shadow states exist, a breakdown of the Fermi liquid theory
and a violation of the Luttinger theorem.

%
%
%

 \end{multicols}

\begin{thebibliography}{99}
\bibitem{FN1}
Note that for the doping concentrations of our 
calculation this shrunk FS   is 
 still large in the sense that 
it does not contains disconnected pieces.
\end{thebibliography}
\end{document}